\documentclass[10pt]{article}
\usepackage{graphicx}
\usepackage{amsmath}
\usepackage{amssymb}
\usepackage{caption2}
\begin{document}
\begin{center}
{\large {\bf \sc{ The  magnetic moment of the  $Z_c(3900)$ as an  axialvector tetraquark state with QCD sum rules
  }}} \\[2mm]
Zhi-Gang  Wang \footnote{E-mail: zgwang@aliyun.com.  }  \\
 Department of Physics, North China Electric Power University, Baoding 071003, P. R. China
\end{center}

\begin{abstract}
In this article, we assign the $Z_c^\pm(3900)$ to be  the diquark-antidiquark  type axialvector  tetraquark  state,   study its magnetic moment with the QCD sum rules in the external weak electromagnetic field by carrying out the operator product expansion up to the vacuum condensates of dimension 8.    We pay  special attention  to matching the hadron side  with the QCD side of the correlation function to obtain solid duality, the routine can be applied to study other electromagnetic properties of the  exotic particles.
\end{abstract}



\section{Introduction}

In 2013, the BESIII collaboration studied  the process  $e^+e^- \to \pi^+\pi^-J/\psi$ at a center-of-mass energy of 4.260 GeV, and observed a structure $Z_c(3900)$ in the $\pi^\pm J/\psi$ invariant mass distribution  \cite{BES3900}. Then the  $Z_c(3900)$ was confirmed by the Belle and CLEO collaborations \cite{Belle3900,CLEO3900}.
  Recently, the BESIII collaboration determined   the  spin and parity of the $Z_c^\pm(3900)$ state to  be $J^P = 1^+$ with a statistical
significance more than $7\sigma$ over other quantum numbers in a partial wave analysis of the process
$e^+e^- \to \pi^+\pi^-J/\psi$ \cite{JP-BES-Zc3900}. If it is really a resonance, the   assignments of  tetraquark state \cite{Maiani1303,Tetraquark3900,WangHuangTao-3900,Nielsen3900} and molecular state \cite{Molecular3900,WangMolecule-3900} are robust according to the non-zero electric charge.  The newly observed exotic  states $X$, $Y$, $Z$, which are excellent candidates for the multiquark states,  provide a good
platform for studying the nonperturbative behavior of QCD and attract many interesting works  of the particle physicists.

The QCD sum rules is a powerful nonperturbative  tool in studying the ground state hadrons, and has  given many successful descriptions of
  the hadronic parameters on the phenomenological side \cite{SVZ79,Reinders85}.  As far as the $Z_c(3900)$ is concerned, the mass and decay width of the $Z_c(3900)$ have been studied with the QCD sum rules in details \cite{WangHuangTao-3900,Nielsen3900,WangMolecule-3900,Azizi-Zc3900-decay,WangZc3900Decay}, while the magnetic moment of the $Z_c(3900)$ is only studied with the light-cone QCD sum rules \cite{LCSR-M-3900}.
The magnetic moment of the $Z_c(3900)$ is a
fundamental parameter as its mass and width, which also has copious information about the
underlying quark structure, can be used to distinguish the preferred configuration
from various theoretical models and deepen our understanding of the underlying
dynamics.

For example, although the proton and neutron have degenerated mass in the isospin limit,
their electromagnetic properties are quite different. If we take them as point particles,
the magnetic moments are $\mu_p=1$ and $\mu_n=0$ in unit of the nucleon magneton from Dirac's theory of relativistic fermions. In 1933, Otto Stern measured the magnetic
moment of the proton, which deviates from one significatively and indicates
that the proton has under-structure. The neutron's anomalous magnetic moment originates from the Pauli form-factor.
So it is interesting to study the interaction of the $Z_c(3900)$ with the photon, which  plays an important role in understanding  its nature. In fact, the works on the magnetic moments of the exotic particles $X$, $Y$, $Z$ are few, only the magnetic moments of the $Z_c(3900)$, $X(5568)$ and $Z_b(10610)$ are studied with the light-cone QCD sum rules \cite{LCSR-M-3900,LCSR-MM-XYZ}.

In this article, we tentatively assign the $Z_c(3900)$ to be the diquark-antidiquark type tetraquark state  and study its magnetic moment with the
QCD sum rules in the external weak electromagnetic field. In Ref.\cite{Ioffe1984}, Ioffe and Smilga (also in Ref.\cite{Balitsky1983}, Balitsky and  Yung) introduce
 a static electromagnetic field which couples to the quarks and polarizes the
QCD vacuum, and extract the nucleon magnetic moments   from linear response to the external electromagnetic field with the QCD sum rules. This approach has been applied successfully to study the magnetic moments of the octet baryons and decuplet baryons \cite{Magnetic-Chiu,Magnetic-Pasupathy-Formula-Value,Magnetic-Chiu-Value,Magnetic-Lee-Formula}. Now we extend this approach to study the magnetic moment of the hidden-charm tetraquark state $Z_c(3900)$.

The article is arranged as follows:  we derive the QCD sum rules for the magnetic moment $\mu$ of the $Z_c(3900)$ in section 2; in Sect.3, we present the numerical results and discussions; and Sect.4 is reserved for our
conclusion.

\section{The magnetic moment   of the $Z_c(3900)$   as an axialvector tetraquark state }

In the following, we write down  the two-point correlation function  $\Pi_{\mu\nu}(p,q)$  in the external electromagnetic field $F$,
\begin{eqnarray}
\Pi_{\mu\nu}(p,q)&=&i\int d^4x e^{ip \cdot x} \langle0|T\left\{J_\mu(x)J_\nu^{\dagger}(0)\right\}|0\rangle_F \, , \\
J_\mu(x)&=&\frac{\varepsilon^{ijk}\varepsilon^{imn}}{\sqrt{2}}\left\{u^j(x)C\gamma_5c^k(x) \bar{d}^m(x)\gamma_\mu C \bar{c}^n(x)-u^j(x)C\gamma_\mu c^k(x)\bar{d}^m(x)\gamma_5C \bar{c}^n(x) \right\} \, ,
\end{eqnarray}
 the $i$, $j$, $k$, $m$, $n$ are color indexes, the $C$ is the charge conjugation matrix.

The correlation function $\Pi_{\mu\nu}(p,q)$ can be rewritten as
\begin{eqnarray}
\Pi_{\mu\nu}(p,q)&=&i\int d^4x e^{ip \cdot x} \langle0|T\left\{J_\mu(x)J_\nu^{\dagger}(0)\right\}|0\rangle\nonumber\\
&&+i\int d^4x e^{ip \cdot x} \langle0|T\left\{J_\mu(x)\left[-ie \int d^4y \,\eta_\alpha(y)\,A^\alpha(y)  \right]J_\nu^{\dagger}(0)\right\}|0\rangle+\cdots \, ,
\end{eqnarray}
where the $\eta_\alpha(y)$ is the electromagnetic current and the $A_\alpha(y)$ is the electromagnetic field.

We can insert  a complete set of intermediate hadronic states with
the same quantum numbers as the current operator $J_\mu(x)$ into the
correlation function $\Pi_{\mu\nu}(p,q)$  to obtain the hadronic representation
\cite{SVZ79,Reinders85}. After isolating the ground state and the first radial excited state
contributions from the pole terms,  we obtain
\begin{eqnarray}
\Pi_{\mu\nu}(p,q)&=&\frac{\lambda_{Z}^2}{M_{Z}^2-p^2} \zeta_\mu(p)\zeta_\nu^*(p)+\frac{\lambda_{Z^\prime}^2}{M_{Z^\prime}^2-p^2} \zeta_\mu(p)\zeta_\nu^*(p) \nonumber\\
&&-\lambda_{Z}^2 e \, \varepsilon^*_\alpha(q)\,\zeta_\mu(p)\,\zeta^*_\nu(p^\prime)\frac{1}{M_{Z}^2-p^2}\langle Z_c(p)|\eta^\alpha(0)|Z_c(p^\prime)\rangle)\frac{1}{M_{Z}^2-p^{\prime2}}   \nonumber\\
&&-\lambda_{Z} \lambda_{Z^\prime}e \, \varepsilon^*_\alpha(q)\,\zeta_\mu(p)\,\zeta^*_\nu(p^\prime)\frac{1}{M_{Z^\prime}^2-p^2}\langle Z_c^\prime(p)|\eta^\alpha(0)|Z_c(p^\prime)\rangle)\frac{1}{M_{Z}^2-p^{\prime2}}   \nonumber\\
&&-\lambda_{Z} \lambda_{Z^\prime}e \, \varepsilon^*_\alpha(q)\,\zeta_\mu(p)\,\zeta^*_\nu(p^\prime)\frac{1}{M_{Z}^2-p^2}\langle Z_c(p)|\eta^\alpha(0)|Z_c^\prime(p^\prime)\rangle)\frac{1}{M_{Z^\prime}^2-p^{\prime2}}  +\cdots \nonumber\\
&=&\frac{\lambda_{Z}^2}{M_{Z}^2-p^2} \left( -g_{\mu\nu} \right)+\frac{\lambda_{Z^\prime}^2}{M_{Z^\prime}^2-p^2} \left( -g_{\mu\nu} \right)+\frac{\lambda_{Z}^2}{\left(M_{Z}^2-p^2\right)\left(M_{Z}^2-p^{\prime2}\right)} \,G_2(0)\,ie\,F_{\mu\nu}\nonumber\\
&&+\frac{\lambda_{Z^\prime}\lambda_{Z}}{\left(M_{Z^\prime}^2-p^2\right)\left(M_{Z}^2-p^{\prime2}\right)} \,\widetilde{G}_2(0)\,ie\,F_{\mu\nu}+\frac{\lambda_{Z}\lambda_{Z^\prime}}{\left(M_{Z}^2-p^2\right)\left(M_{Z^\prime}^2-p^{\prime2}\right)} \,\widetilde{G}_2(0)\,ie\,F_{\mu\nu}+\cdots\nonumber\\
&=&\Pi(p^{\prime2},p^2)i\,e\,F_{\mu\nu}+\cdots\, ,
\end{eqnarray}
where $F_{\mu\nu}=\partial_\mu A_\nu-\partial_\nu A_\mu$, the pole residue (or coupling) $\lambda_{Z^{(\prime)}}$ is defined by
\begin{eqnarray}
 \langle 0|J_\mu(0)|Z^{(\prime)}(p)\rangle=\lambda_{Z^{(\prime)}} \, \zeta_\mu(p) \, ,
\end{eqnarray}
the $\varepsilon_\alpha$ and $\zeta_\mu$ are the polarization vectors of the photon and the axial-vector meson  $Z_c^{(\prime)}$, respectively.
The hadronic matrix element $\langle Z_c(p)|\eta_\alpha(0)|Z_c(p^\prime)\rangle$ can be parameterized by three form-factors,
\begin{eqnarray}
\langle Z_c(p)|\eta_\alpha(0)|Z_c(p^\prime)\rangle &=&G_1(Q^2)\,\zeta^*(p)\cdot \zeta(p^\prime)\left(p_\alpha+p^\prime_\alpha \right)+G_2(Q^2)\left[\zeta_\alpha(p^\prime)\zeta^*(p)\cdot q-\zeta^*_\alpha(p)\zeta(p^\prime)\cdot q \right] \nonumber\\
&&-G_3(Q^2)\frac{1}{2M_{Z}^2}\zeta^*(p)\cdot q \, \zeta(p^\prime)\cdot q \left(p_\alpha+p^\prime_\alpha \right)\, ,
\end{eqnarray}
with $Q^2=-q^2$, $p^\prime=p+q$ \cite{Brodsky1992}.
The Lorentz-invariant form-factors $G_1(Q^2)$, $G_2(Q^2)$ and $G_3(Q^2)$ are related to the charge, magnetic
and quadrupole form-factors,
\begin{eqnarray}
G_C&=&G_1+\frac{2}{3}\eta G_Q\, ,\nonumber\\
G_M&=&-G_2\, ,\nonumber\\
G_Q&=&G_1+G_2+(1+\eta)G_3\, ,
\end{eqnarray}
respectively, where $\eta=\frac{Q^2}{4M_{Z}^2}$ is a kinematic factor. At zero momentum transfer, these form-factors
are proportional to the usual static quantities of the charge $e$, magnetic moment $\mu_Z$ and
quadrupole moment $Q_1$,
\begin{eqnarray}
e\, G_C(0)&=&e\, ,\nonumber\\
e\, G_M(0)&=&2M_{Z}\,\mu_Z\, ,\nonumber\\
e\, G_Q(0)&=&M_{Z}^2\,Q_1\, .
\end{eqnarray}
The  hadronic matrix elements $\langle Z_c(p)|\eta_\alpha(0)|Z^\prime_c(p^\prime)\rangle$ and $\langle Z_c^\prime(p)|\eta_\alpha(0)|Z_c(p^\prime)\rangle$ are parameterized analogously by the form-factors $\widetilde{G}_1(Q^2)$, $\widetilde{G}_2(Q^2)$ and $\widetilde{G}_3(Q^2)$. In this article, we choose the tensor structure $F_{\mu\nu}$ for analysis.

The   current $J_\mu(x)$  also has non-vanishing couplings  with the scattering states  $D
D^\ast$, $J/\psi \pi$, $J/\psi \rho$, etc  \cite{PDG}.
 In the following, we
study the contributions of the  intermediate   meson-loops to the correlation function $\Pi_{\mu\nu}(p,q)$,
\begin{eqnarray}
\Pi_{\mu\nu}(p,q)&=&\frac{\widehat{\lambda}_{Z}^2}{\widehat{M}_{Z}^2+\Sigma(p) -p^2} \left( -g_{\mu\nu} \right)+\frac{\lambda_{\widehat{Z}^\prime}^2}{\widehat{M}_{Z^\prime}^2+\Sigma(p)-p^2} \left( -g_{\mu\nu} \right)\nonumber\\
&&+\frac{\widehat{\lambda}_{Z}^2}{\left(\widehat{M}_{Z}^2+\Sigma(p)-p^2\right)\left(\widehat{M}_{Z}^2+\Sigma(p^\prime)-p^{\prime2}\right)} \,G_2(0)\,ie\,F_{\mu\nu}\nonumber\\
&&+\frac{\widehat{\lambda}_{Z^\prime}\widehat{\lambda}_{Z}}{\left(\widehat{M}_{Z^\prime}^2+\Sigma(p)-p^2\right)\left(\widehat{M}_{Z}^2+\Sigma(p^\prime)-p^{\prime2}\right)} \,\widetilde{G}_2(0)\,ie\,F_{\mu\nu}\nonumber\\
&&+\frac{\widehat{\lambda}_{Z}\widehat{\lambda}_{Z^\prime}}{\left(\widehat{M}_{Z}^2+\Sigma(p)-p^2\right)\left(\widehat{M}_{Z^\prime}^2+\Sigma(p^\prime)-p^{\prime2}\right)} \,\widetilde{G}_2(0)\,ie\,F_{\mu\nu}+\cdots  \, ,
\end{eqnarray}
where the self-energy
$\Sigma(p)=\Sigma_{DD^*}(p)+\Sigma_{J/\psi\pi}(p)+\Sigma_{J/\psi\rho}(p)+\cdots$.
The $\widehat{\lambda}_{Z/Z^\prime}$ and $\widehat{M}_{Z/Z^\prime}$ are bare quantities to absorb the divergences in the self-energies $\Sigma(p)$ and $\Sigma(p^\prime)$.
The renormalized self-energies  contribute  a finite imaginary part to modify the dispersion relation,
\begin{eqnarray}
\Pi_{\mu\nu}(p,q)&=&\frac{\lambda_{Z}^2}{M_{Z}^2-i\sqrt{p^2}\,\Gamma(p^2) -p^2} \left( -g_{\mu\nu} \right)+\frac{\lambda_{Z^\prime}^2}{M_{Z^\prime}^2-i\sqrt{p^2}\,\Gamma^\prime(p^2)-p^2} \left( -g_{\mu\nu} \right)\nonumber\\
&&+\frac{\lambda_{Z}^2}{\left(M_{Z}^2-i\sqrt{p^2}\,\Gamma(p^2)-p^2\right)\left(M_{Z}^2-i\sqrt{p^{\prime2}}\,\Gamma(p^{\prime2})-p^{\prime2}\right)} \,G_2(0)\,ie\,F_{\mu\nu}\nonumber\\
&&+\frac{\lambda_{Z^\prime}\lambda_{Z}}{\left(M_{Z^\prime}^2-i\sqrt{p^2}\,\Gamma^\prime(p^2)-p^2\right)\left(M_{Z}^2
-i\sqrt{p^{\prime2}}\,\Gamma(p^{\prime2})-p^{\prime2}\right)} \,\widetilde{G}_2(0)\,ie\,F_{\mu\nu}\nonumber\\
&&+\frac{\lambda_{Z}\lambda_{Z^\prime}}{\left(M_{Z}^2-i\sqrt{p^2}\,\Gamma(p^2)-p^2\right)\left(M_{Z^\prime}^2
-i\sqrt{p^{\prime2}}\,\Gamma^\prime(p^{\prime2})-p^{\prime2}\right)} \,\widetilde{G}_2(0)\,ie\,F_{\mu\nu}+\cdots  \, ,
\end{eqnarray}
the physical  width of the ground state $\Gamma(M_Z^2)=\Gamma_{Z_c(3900)}=(28.1\pm 2.6)\, \rm{MeV}$ is small enough \cite{PDG}, while the effect of the large width of the first radial excited  state $\Gamma^\prime(M^2_{Z^\prime})=\Gamma_{Z(4430)}=181\pm 31\,\rm{MeV}$ can be absorbed into the continuum states (the $Z(4430)$ can be tentatively assigned to be the first radial excitation of the $Z_c(3900)$ based on the QCD sum rules \cite{Wang4430}),
 the zero width approximation in  the hadronic spectral density works, the contaminations of the intermediate meson-loops are expected
 to be small and can be neglected safely \cite{Wang-EPJC-Scalar-meson}. In fact, even the effect of the  large width of the $Z_c(4200)$ ($\Gamma_{Z_c(4200)}=370\pm70{}^{+ 70}_{-132}\,\rm{MeV}$) can be safely absorbed into the pole residue \cite{Wang-Zc4200}. For detailed discussions about this subject, one can consult Refs.\cite{WangHuangTao-3900,Wang-Zc4200}.

 In the following,  we briefly outline  the operator product expansion for the correlation function $\Pi_{\mu\nu}(p,q)$  in perturbative
QCD.  We contract the quark fields in the correlation function
$\Pi_{\mu\nu}(p,q)$ with Wick theorem, obtain the result,
\begin{eqnarray}
\Pi_{\mu\nu}(p,q)&=&-\frac{i\varepsilon^{ijk}\varepsilon^{imn}\varepsilon^{i^{\prime}j^{\prime}k^{\prime}}\varepsilon^{i^{\prime}m^{\prime}n^{\prime}}}{2}\int d^4x e^{ip \cdot x}   \nonumber\\
&&\left\{{\rm Tr}\left[ \gamma_5C^{kk^{\prime}}(x)\gamma_5 CU^{jj^{\prime}T}(x)C\right] {\rm Tr}\left[ \gamma_\nu C^{n^{\prime}n}(-x)\gamma_\mu C D^{m^{\prime}mT}(-x)C\right] \right. \nonumber\\
&&+{\rm Tr}\left[ \gamma_\mu C^{kk^{\prime}}(x)\gamma_\nu CU^{jj^{\prime}T}(x)C\right] {\rm Tr}\left[ \gamma_5 C^{n^{\prime}n}(-x)\gamma_5 C D^{m^{\prime}mT}(-x)C\right] \nonumber\\
&&+{\rm Tr}\left[ \gamma_\mu C^{kk^{\prime}}(x)\gamma_5 CU^{jj^{\prime}T}(x)C\right] {\rm Tr}\left[ \gamma_\nu C^{n^{\prime}n}(-x)\gamma_5 C D^{m^{\prime}mT}(-x)C\right] \nonumber\\
 &&\left.+{\rm Tr}\left[ \gamma_5 C^{kk^{\prime}}(x)\gamma_\nu CU^{jj^{\prime}T}(x)C\right] {\rm Tr}\left[ \gamma_5 C^{n^{\prime}n}(-x)\gamma_\mu C D^{m^{\prime}mT}(-x)C\right] \right\} \nonumber\\
 &=&\Pi(p^{\prime2},p^2)\,i\,e\,F_{\mu\nu}+\cdots \, ,
\end{eqnarray}
where  the $U_{ij}(x)$, $D_{ij}(x)$ and $C_{ij}(x)$ are the full $u$, $d$ and $c$ quark propagators, respectively (the $U_{ij}(x)$ and $D_{ij}(x)$ can be written as $S_{ij}(x)$ for simplicity),
\begin{eqnarray}
S_{ij}(x)&=& \frac{i\delta_{ij}\!\not\!{x}}{ 2\pi^2x^4}-\frac{\delta_{ij}\langle
\bar{q}q\rangle}{12} -\frac{\delta_{ij}x^2\langle \bar{q}g_s\sigma Gq\rangle}{192} -\frac{ig_sG^{a}_{\alpha\beta}t^a_{ij}\left(\!\not\!{x}
\sigma^{\alpha\beta}+\sigma^{\alpha\beta} \!\not\!{x}\right)}{32\pi^2x^2}     -\frac{1}{8}\langle\bar{q}_j\sigma^{\mu\nu}q_i \rangle \sigma_{\mu\nu}\nonumber\\
&&+\frac{i \delta_{ij}\, e e_q F_{\alpha\beta}\left(\!\not\!{x}
\sigma^{\alpha\beta}+\sigma^{\alpha\beta} \!\not\!{x}\right)}{32\pi^2x^2} -\frac{\delta_{ij}\langle
\bar{q}q\rangle\,\chi\, \sigma^{\alpha\beta}\, e e_q F_{\alpha\beta}}{24}\nonumber\\
 &&+\frac{\delta_{ij}\langle\bar{q}q\rangle\,e e_q F_{\alpha\beta} }{288}\left(\sigma^{\alpha\beta}x^2-2x_{\lambda}x^{\beta}\sigma^{\lambda\alpha} \right)\nonumber\\
&&+\frac{\delta_{ij}\langle\bar{q}q\rangle\,e e_q F_{\alpha\beta} }{576}\left[ \sigma^{\alpha\beta}x^2\left(\kappa+\xi\right)-2x_\lambda x^\beta \sigma^{\lambda\alpha}\left(\kappa-\frac{\xi}{2} \right)\right]+\cdots\, ,
\end{eqnarray}
\begin{eqnarray}
C_{ij}(x)&=&\frac{i}{(2\pi)^4}\int d^4k e^{-ik \cdot x} \left\{
\frac{\delta_{ij}}{\!\not\!{k}-m_c}
-\frac{g_sG^n_{\alpha\beta}t^n_{ij}}{4}\frac{\sigma^{\alpha\beta}(\!\not\!{k}+m_c)+(\!\not\!{k}+m_c)
\sigma^{\alpha\beta}}{(k^2-m_c^2)^2}\right.\nonumber\\
&&\left. +\frac{\delta_{ij} e e_q F_{\alpha\beta} }{4}\frac{\sigma^{\alpha\beta}(\!\not\!{k}+m_c)+(\!\not\!{k}+m_c)
\sigma^{\alpha\beta}}{(k^2-m_c^2)^2}+\cdots\right\} \, ,
\end{eqnarray}
and  $t^n=\frac{\lambda^n}{2}$, the $\lambda^n$ is the Gell-Mann matrix \cite{WangHuangTao-3900,Reinders85,Ioffe1984,Magnetic-Pasupathy-Formula-Value,Magnetic-Lee-Formula,Pascual-1984}.
We retain the term $\langle\bar{q}_j\sigma_{\mu\nu}q_i \rangle$  originates from  Fierz re-ordering of the $\langle q_i \bar{q}_j\rangle$ to  absorb the gluons  emitted from other quark lines to form $\langle\bar{q}_j g_s G^a_{\alpha\beta} t^a_{mn}\sigma_{\mu\nu} q_i \rangle$   to extract the mixed condensate   $\langle\bar{q}g_s\sigma G q\rangle$ \cite{WangHuangTao-3900}. The new condensates or vacuum susceptibilities $\chi$, $\kappa$ and $\xi$ induced by the external electromagnetic field  are defined by
\begin{eqnarray}
\langle 0|\bar{q}\sigma_{\alpha\beta}q|0\rangle_F&=& e\,e_q\,\chi\, \langle \bar{q}q\rangle\, F_{\alpha\beta}\, , \nonumber\\
\langle 0|\bar{q}g_s G_{\alpha\beta}q|0\rangle_F&=& e\,e_q\,\kappa\, \langle \bar{q}q\rangle\, F_{\alpha\beta}\, , \nonumber\\
\varepsilon_{\alpha\beta\lambda\tau}\langle 0|\bar{q}g_s G^{\lambda\tau}\gamma_5q|0\rangle_F&=&i e\,e_q\,\xi\, \langle \bar{q}q\rangle\, F_{\alpha\beta}\, ,
\end{eqnarray}
with the convention $\varepsilon^{0123}=-\varepsilon_{0123}=1$ \cite{Ioffe1984,Magnetic-Pasupathy-Formula-Value,Magnetic-Lee-Formula}.
Then we compute  the integrals both in the coordinate and momentum spaces,  and obtain the correlation function $\Pi(p^{\prime2},p^2)\,F_{\mu\nu}$ therefore the spectral density  at the level of   quark-gluon degrees  of freedom.   Furthermore, we take into account contributions of the new condensates originate from the quark interacting with
the external electromagnetic field as well as the vacuum gluons,
\begin{eqnarray}
\langle 0| q^i(x)\bar{q}^j(0)G^n_{\alpha\beta}(x)|0\rangle_F&=&-\frac{\langle\bar{q}q\rangle}{16}\,t^n_{ij}\,e e_q \left( \kappa\, F_{\alpha\beta}-\frac{i}{4}\xi\, \gamma_5 \,\varepsilon_{\alpha\beta\lambda\tau}F^{\lambda\tau}\right)\, ,
\end{eqnarray}
see the Feynman diagrams shown in Figs.1-2.

\begin{figure}
 \centering
  \includegraphics[totalheight=4cm,width=12cm]{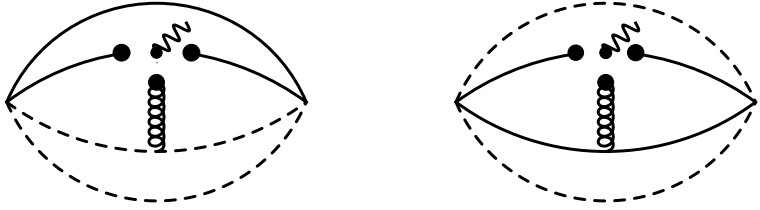}
         \caption{ The contributions  $\langle\bar{q}q\rangle\, F_{\alpha\beta}$ come from the terms     $\langle 0| q^i(x)\bar{q}^j(0)G^n_{\alpha\beta}(x)|0\rangle_F$, where the external lines denote the photon field. Other
diagrams obtained by interchanging of the heavy quark lines (dashed lines) or light quark lines (solid lines) are implied. }
\end{figure}

\begin{figure}
 \centering
  \includegraphics[totalheight=4cm,width=12cm]{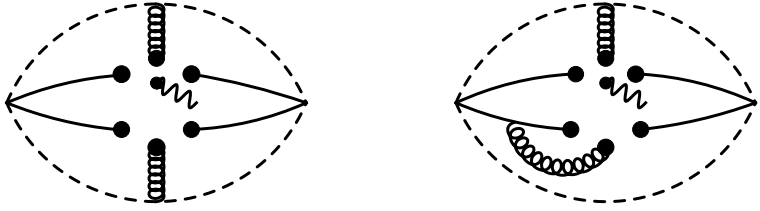}
         \caption{ The contributions  $\langle\bar{q}q\rangle \langle\bar{q}g_s \sigma Gq\rangle\, F_{\alpha\beta}$ come from the terms     $\langle 0| q^i(x)\bar{q}^j(0)G^n_{\alpha\beta}(x)|0\rangle_F$, where the external lines denote the photon field. Other
diagrams obtained by interchanging of the heavy quark lines (dashed lines) or light quark lines (solid lines) are implied.  }
\end{figure}

We have to be cautious in matching the QCD side of the correlation function $\Pi(p^{\prime2},p^2)$ with the hadron  side of the correlation function $\Pi(p^{\prime 2},p^2)$, as there appears  the variable $p^{\prime2}=(p+q)^2$.
We rewrite the correlation function $\Pi_H(p^{\prime 2},p^2)$ on the hadron  side into the following form through dispersion relation,
\begin{eqnarray}
\Pi_{H}(p^{\prime 2},p^2)&=&\int_{4m_c^2}^{s_{Z}^0}ds^\prime \int_{4m_c^2}^{s^0_{Z}}ds  \frac{\rho_H(s^\prime,s)}{(s^\prime-p^{\prime2})(s-p^2)}+\cdots\, ,
\end{eqnarray}
where the $\rho_H(s^\prime,s)$   is the hadronic spectral density,
\begin{eqnarray}
\rho_H(s^\prime,s)&=&{\lim_{\epsilon_2\to 0}}\,\,{\lim_{\epsilon_1\to 0}} \,\,\frac{ {\rm Im}_{s^\prime}\, {\rm Im}_{s}\,\Pi_H(s^\prime+i\epsilon_2,s+i\epsilon_1) }{\pi^2} \nonumber \\
&=&\lambda_Z^2\, G_2(0)\,\delta\left(s^\prime -M_Z^2\right)\delta\left(s -M_Z^2\right)+ \lambda_{Z^\prime}\lambda_Z\, \widetilde{G}_2(0)\,\delta\left(s^\prime -M_{Z^\prime}^2\right)\delta\left(s -M_Z^2\right)\nonumber\\
&&+\lambda_{Z^\prime}\lambda_Z\, \widetilde{G}_2(0)\,\delta\left(s^\prime -M_{Z}^2\right)\delta\left(s -M_{Z^\prime}^2\right)+\cdots\, ,
\end{eqnarray}
the $s_Z^0$ is the continuum threshold parameter, we add the subscript $H$ to denote the hadron side.
However, on the QCD side, the QCD spectral density  $\rho_{QCD}(s^\prime,s)$ does not exist,
\begin{eqnarray}
\rho_{QCD}(s^\prime,s)&=&{\lim_{\epsilon_2\to 0}} \,\,{\lim_{\epsilon_1\to 0}}\,\,\frac{ {\rm Im}_{s^\prime}\, {\rm Im}_{s}\,\Pi_{QCD}(s^\prime+i\epsilon_2,s+i\epsilon_1) }{\pi^2} \nonumber\\
&=&0\, ,
\end{eqnarray}
because
\begin{eqnarray}
{\lim_{\epsilon_2\to 0}}\,\,\frac{ {\rm Im}_{s^\prime}\,\Pi_{QCD}(s^\prime+i\epsilon_2,p^2) }{\pi} &=&0\, ,
\end{eqnarray}
we add the subscript $QCD$ to denote the QCD side.

On the QCD side, the correlation function $\Pi_{QCD}(p^2)$ (the $\Pi_{QCD}(p^{\prime2},p^2)$ is reduced to $\Pi_{QCD}(p^2)$ due to lacking dependence on the variable $p^{\prime2}$) can be written into the following form through dispersion relation,
\begin{eqnarray}
\Pi_{QCD}(p^2)&=&  a_n\, p^{2n}+ \int_{4m_c^2}^{s^0_{Z}}ds    \frac{\rho_{QCD}(s)}{s-p^2}+\cdots\, ,
\end{eqnarray}
where the $\rho_{QCD}(s)$   is the QCD spectral density,
\begin{eqnarray}
\rho_{QCD}(s)&=&   \,\,{\lim_{\epsilon_1\to 0}}\,\,\frac{  {\rm Im}_{s} \,\Pi_{QCD}(s+i\epsilon_1) }{\pi} \, ,
\end{eqnarray}
which is independent on the variable $p^{\prime2}$, the coefficients $a_n$ with $n=0,\,1,\,2, \cdots$ are some constants, the terms $a_n\, p^{2n}$ disappear after performing the Borel transform with respect to the variable $P^2=-p^2$, we can neglect the terms $a_n\, p^{2n}$ safely.

We math the hadron side   with the QCD side of the correlation function, and carry out the integral over $ds^\prime$  firstly to obtain the solid duality,
 \begin{eqnarray}
  \int_{4m_c^2}^{s^0_{Z}}ds  \frac{\rho_{QCD}(s)}{s-p^2}&=& \int_{4m_c^2}^{s^0_{Z}}ds \frac{1}{s-p^2} \left[\int_{4m_c^2}^{\infty}ds^\prime \frac{\rho_H(s^\prime,s)}{s^\prime-p^{\prime2}} \right]\nonumber\\
  &=& \frac{\lambda_{Z}^2\,G_2(0)}{\left(M_{Z}^2-p^2\right)\left(M_{Z}^2-p^{\prime2}\right)}+  \int_{4m_c^2}^{s^0_{Z}}ds \frac{1}{s-p^2} \left[\int_{s^0_Z}^{\infty}ds^\prime \frac{\rho_H(s^\prime,s)}{s^\prime-p^{\prime2}} \right]\nonumber\\
 &=& \frac{\lambda_{Z}^2\,G_2(0)}{\left(M_{Z}^2-p^2\right)\left(M_{Z}^2-p^{\prime2}\right)}+  \frac{\lambda_Z^2\, C}{M_Z^2-p^2}\, ,
\end{eqnarray}
where we introduce the unknown parameter $C$ to denote the transition between the ground state and the excited states,
\begin{eqnarray}
\lambda_Z^2\, C&=&  \int_{4m_c^2}^{s^0_{Z}}ds \int_{s^0_Z}^{\infty}ds^\prime \frac{\rho_H(s^\prime,s)}{s^\prime-p^{\prime2}}=\frac{\lambda_{Z^\prime}\lambda_Z\, \widetilde{G}_2(0)}{M_{Z^\prime}^2-p^{\prime2}}+\cdots\, .
\end{eqnarray}
In numerical calculation,   we smear    dependency of the  $C$  on the momentum $p^{\prime2}$ and take it  as a free parameter, and choose the suitable value  to
eliminate the contaminations from the high resonances and continuum states to obtain the stable QCD sum rule with respect to the variation of
the Borel parameter $T^2$.

Now  we set $p^{\prime2}=p^2$ and perform the Borel transform  with respect to
the variable $P^2=-p^2$ to obtain  the following QCD sum rule:
\begin{eqnarray}
\left[\frac{G_2(0)}{T^2}+C\right]\, \lambda_{Z}^2\, \exp\left(-\frac{M_{Z}^2}{T^2}\right)= \int_{4m_c^2}^{s_0} ds\, \rho_{QCD}(s) \, \exp\left(-\frac{s}{T^2}\right) \, ,
\end{eqnarray}
where
\begin{eqnarray}
\rho_{QCD}(s)&=&\rho_0(s)+\rho_3(s)+\rho_5(s)+\rho_6(s)+\rho_8(s)\, ,
\end{eqnarray}

\begin{eqnarray}
\rho_{0}(s)&=&\frac{1}{256\pi^6}\int_{y_i}^{y_f}dy \int_{z_i}^{1-y}dz \, yz\, (1-y-z)\left(s-\overline{m}_c^2\right)^2\left(4s-\overline{m}_c^2 \right)  \, ,
\end{eqnarray}

\begin{eqnarray}
\rho_{3}(s)&=&-\frac{m_c\langle \bar{q}q\rangle}{96\pi^4}\int_{y_i}^{y_f}dy \int_{z_i}^{1-y}dz \, (y+z)\, s \nonumber\\
&&+\frac{m_c\langle \bar{q}q\rangle \,\chi}{64\pi^4}\int_{y_i}^{y_f}dy \int_{z_i}^{1-y}dz \, (y+z)\,(1-y-z) \left(s-\overline{m}_c^2\right)\left(2s-\overline{m}_c^2 \right) \nonumber\\
&&+\frac{m_c\langle \bar{q}q\rangle \,\kappa}{384\pi^4}\int_{y_i}^{y_f}dy \int_{z_i}^{1-y}dz \, (y+z)\,  \left(4s-3\overline{m}_c^2\right)  \nonumber\\
&&+\frac{m_c\langle \bar{q}q\rangle \,\xi}{768\pi^4}\int_{y_i}^{y_f}dy \int_{z_i}^{1-y}dz \, (y+z)\,  \left(5s-3\overline{m}_c^2\right)  \nonumber\\
&&-\frac{m_c\langle \bar{q}q\rangle \,(2\kappa+\xi)}{256\pi^4}\int_{y_i}^{y_f}dy \int_{z_i}^{1-y}dz \, \left(\frac{y}{z}+\frac{z}{y}\right)\,(1-y-z)  \left(3s-2\overline{m}_c^2\right)  \nonumber\\
&&-\frac{m_c\langle \bar{q}q\rangle \,\kappa}{256\pi^4}\int_{y_i}^{y_f}dy \int_{z_i}^{1-y}dz \,(2-y-z)    \left(2s-\overline{m}_c^2\right)  \nonumber\\
&&+\frac{m_c\langle \bar{q}q\rangle \,\xi}{512\pi^4}\int_{y_i}^{y_f}dy \int_{z_i}^{1-y}dz \, \left[\left(\frac{y}{z}+\frac{z}{y}\right)\,(1-y-z) -(y+z)\right] \left(s-\overline{m}_c^2\right)  \, ,
\end{eqnarray}

\begin{eqnarray}
\rho_{5}(s)&=&\frac{m_c\langle \bar{q}g_s\sigma Gq\rangle}{256\pi^4}\int_{y_i}^{y_f}dy   \, \left[1+s\,\delta\left(s-\widetilde{m}_c^2 \right) \right] \nonumber\\
&&-\frac{m_c\langle \bar{q}g_s\sigma Gq\rangle}{1536\pi^4}\int_{y_i}^{y_f}dy  \int_{z_i}^{1-y}dz \, \left[1+s\,\delta\left(s-\overline{m}_c^2 \right) \right] \nonumber\\
&&-\frac{5m_c\langle \bar{q}g_s\sigma Gq\rangle}{1536\pi^4}\int_{y_i}^{y_f}dy \int_{z_i}^{1-y}dz   \,\left(\frac{z}{y}+\frac{y}{z} \right) \left[1+s\,\delta\left(s-\overline{m}_c^2 \right) \right] \, ,
\end{eqnarray}

\begin{eqnarray}
\rho_{6}(s)&=&-\frac{m_c^2\langle \bar{q}q\rangle^2\,\chi}{24\pi^2}\int_{y_i}^{y_f}dy   -\frac{\langle \bar{q}q\rangle^2\,\xi}{384\pi^2}\int_{y_i}^{y_f}dy  \nonumber\\
&&-\frac{m_c^2\langle \bar{q}q\rangle^2\,(\kappa+2)}{72\pi^2}\int_{y_i}^{y_f}dy\, \left(1+\frac{s}{2T^2} \right) \delta\left(s-\widetilde{m}_c^2 \right)  \nonumber\\
&&-\frac{m_c^2\langle \bar{q}q\rangle^2\,\xi}{288\pi^2}\int_{y_i}^{y_f}dy\, \left(1+\frac{2s}{T^2} \right) \delta\left(s-\widetilde{m}_c^2 \right)  \nonumber\\
&&+\frac{\langle \bar{q}q\rangle^2\,(3\kappa+\xi)}{192\pi^2}\int_{y_i}^{y_f}dy\, s\, \delta\left(s-\widetilde{m}_c^2 \right)  \, ,
\end{eqnarray}

\begin{eqnarray}
\rho_{8}(s)&=&\frac{m_c^2\langle \bar{q}q\rangle\langle \bar{q}g_s\sigma Gq\rangle \,\chi}{96\pi^2}\int_{y_i}^{y_f}dy \, \left(1+\frac{s}{T^2} \right) \delta\left(s-\widetilde{m}_c^2 \right)   \nonumber\\
&&+\frac{m_c^2\langle \bar{q}q\rangle\langle \bar{q}g_s\sigma Gq\rangle \,(\kappa+2)}{576\pi^2T^2}\int_{y_i}^{y_f}dy \, \left(1+\frac{s}{T^2}+\frac{s^2}{T^4}  \right) \delta\left(s-\widetilde{m}_c^2 \right)   \nonumber\\
&&-\frac{m_c^2\langle \bar{q}q\rangle\langle \bar{q}g_s\sigma Gq\rangle\, \xi}{1152\pi^2T^2}\int_{y_i}^{y_f}dy \, \left(1+\frac{s}{T^2}-\frac{2s^2}{T^4}  \right) \delta\left(s-\widetilde{m}_c^2 \right)   \nonumber\\
&&-\frac{ 13\langle \bar{q}q\rangle\langle \bar{q}g_s\sigma Gq\rangle \,\chi}{2304\pi^2}\int_{y_i}^{y_f}dy \, s\, \delta\left(s-\widetilde{m}_c^2 \right)   \nonumber\\
&&-\frac{13\langle \bar{q}q\rangle\langle \bar{q}g_s\sigma Gq\rangle\, (\kappa+2)}{13824\pi^2}\int_{y_i}^{y_f}dy \, \left(\frac{s}{T^2}+\frac{s^2}{T^4}  \right) \delta\left(s-\widetilde{m}_c^2 \right)   \nonumber\\
&&+\frac{13\langle \bar{q}q\rangle\langle \bar{q}g_s\sigma Gq\rangle\, \xi}{27648\pi^2}\int_{y_i}^{y_f}dy \, \left(\frac{s}{T^2}-\frac{2s^2}{T^4}  \right) \delta\left(s-\widetilde{m}_c^2 \right)   \nonumber\\
&&+\frac{ \langle \bar{q}q\rangle\langle \bar{q}g_s\sigma Gq\rangle \,\chi}{768\pi^2}\int_{y_i}^{y_f}dy \,\left[1+ \frac{2s}{3}\, \delta\left(s-\widetilde{m}_c^2 \right) \right]  \nonumber\\
&&+\frac{ \langle \bar{q}q\rangle\langle \bar{q}g_s\sigma Gq\rangle \,(\kappa+2)}{3456\pi^2}\int_{y_i}^{y_f}dy \, \left(1+\frac{s}{T^2}+\frac{s^2}{2T^4}  \right) \delta\left(s-\widetilde{m}_c^2 \right)   \nonumber\\
&&+\frac{ \langle \bar{q}q\rangle\langle \bar{q}g_s\sigma Gq\rangle \,\xi}{13824\pi^2}\int_{y_i}^{y_f}dy \, \left(1+\frac{5s}{2T^2}+\frac{2s^2}{T^4}  \right) \delta\left(s-\widetilde{m}_c^2 \right)   \nonumber\\
&&-\frac{ \langle \bar{q}q\rangle\langle \bar{q}g_s\sigma Gq\rangle \,(3\kappa+\xi)}{768\pi^2}\int_{y_i}^{y_f}dy \, \frac{s^2}{T^4} \, \delta\left(s-\widetilde{m}_c^2 \right)   \nonumber\\
&&+\frac{ \langle \bar{q}q\rangle\langle \bar{q}g_s\sigma Gq\rangle \,\xi}{768\pi^2}\int_{y_i}^{y_f}dy \, \left(1+\frac{s}{2T^2}  \right) \delta\left(s-\widetilde{m}_c^2 \right)   \nonumber\\
&&+\frac{59 \langle \bar{q}q\rangle\langle \bar{q}g_s\sigma Gq\rangle \,(2\kappa+\xi)}{18432\pi^2}\int_{y_i}^{y_f}dy \, \frac{s}{T^2} \, \delta\left(s-\widetilde{m}_c^2 \right)   \nonumber\\
&&-\frac{11 \langle \bar{q}q\rangle\langle \bar{q}g_s\sigma Gq\rangle \,\kappa}{4608\pi^2}\int_{y_i}^{y_f}dy \, \frac{s}{T^2} \, \delta\left(s-\widetilde{m}_c^2 \right)  \, ,
\end{eqnarray}
 $y_{f}=\frac{1+\sqrt{1-4m_c^2/s}}{2}$,
$y_{i}=\frac{1-\sqrt{1-4m_c^2/s}}{2}$, $z_{i}=\frac{y
m_c^2}{y s -m_c^2}$, $\overline{m}_c^2=\frac{(y+z)m_c^2}{yz}$,
$ \widetilde{m}_c^2=\frac{m_c^2}{y(1-y)}$, $\int_{y_i}^{y_f}dy \to \int_{0}^{1}dy$, $\int_{z_i}^{1-y}dz \to \int_{0}^{1-y}dz$ when the $\delta$ functions $\delta\left(s-\overline{m}_c^2\right)$ and $\delta\left(s-\widetilde{m}_c^2\right)$ appear, we neglect the small contributions of the gluon condensate.

According to Eqs.(22-23), the non-diagonal transitions between the ground state and the excited states can be written as
\begin{eqnarray}
\frac{\lambda_Z^2\, C}{M_Z^2-p^2}&=&\frac{\lambda_{Z^\prime}\lambda_Z\, \widetilde{G}_2(0)}{\left(M_Z^2-p^2\right)\left(M_{Z^\prime}^2-p^{\prime2}\right)}+\cdots\, .
\end{eqnarray}
If we set $p^{\prime2}=p^2$ and perform the Borel transform  with respect to
the variable $P^2=-p^2$, we can obtain
\begin{eqnarray}
\frac{\lambda_{Z^\prime}\lambda_Z\, \widetilde{G}_2(0)}{M_{Z^\prime}^2-M_Z^2}\left[ \exp\left(-\frac{M_{Z}^2}{T^2}\right)-\exp\left(-\frac{M_{Z^\prime}^2}{T^2}\right)\right] +\cdots\, .
\end{eqnarray}
In Ref.\cite{Wang4430}, we observe that the $Z(4430)$ can be tentatively assigned to be the first radial excitation of the $Z_c(3900)$ based on the QCD sum rules.
If we set $Z^\prime=Z(4430)$, then 
\begin{eqnarray}
\frac{\exp\left(-\frac{M_{Z^\prime}^2}{T^2}\right)}{\exp\left(-\frac{M_{Z}^2}{T^2}\right)}&=&0.11\sim0.17\, ,
\end{eqnarray}
for the Borel parameter $T^2=(2.2-2.8)\,\rm{GeV}^2$. So the terms $\exp\left(-\frac{M_{Z^\prime}^2}{T^2}\right)$, $\exp\left(-\frac{M_{Z^{\prime\prime}}^2}{T^2}\right)$, $\cdots$ can be neglected approximately,
the  non-diagonal transitions can be approximated as $\lambda_Z^2 C \exp\left(-\frac{M_{Z}^2}{T^2}\right)$.

\section{Numerical results and discussions}	
The input parameters on the QCD side are taken to be the standard values
$\langle\bar{q}q \rangle=-(0.24\pm 0.01\, \rm{GeV})^3$,
 $\langle\bar{q}g_s\sigma G q \rangle=m_0^2\langle \bar{q}q \rangle$,
$m_0^2=(0.8 \pm 0.1)\,\rm{GeV}^2$,
 at the energy scale  $\mu=1\, \rm{GeV}$
\cite{SVZ79,Reinders85,Magnetic-Pasupathy-Formula-Value,Magnetic-Chiu-Value,Colangelo-Review}, and $m_{c}(m_c)=(1.28\pm0.03)\,\rm{GeV}$
 from the Particle Data Group \cite{PDG}. Furthermore, we set $m_u=m_d=0$ due to the small current quark masses.
 For the new condensates or vacuum susceptibilities $\chi$, $\kappa$ and $\xi$ induced by the external electromagnetic field, we take two set parameters:\\
 {\bf (I)} the old values $ \chi=-3\,\rm{GeV}^{-2}$, $\kappa=-0.75$, $\xi=1.5$
 at the energy scale  $\mu=1\, \rm{GeV}$
  fitted  in the QCD sum rules for the magnetic moments of the $p$, $n$ and $\Lambda$ \cite{Magnetic-Pasupathy-Formula-Value,Magnetic-Chiu-Value};\\
 {\bf (II)} the new values $ \chi=-(3.15\pm 0.30)\,\rm{GeV}^{-2}$, $\kappa=-0.2$, $\xi=0.4$
 at the energy scale  $\mu=1\, \rm{GeV}$
  determined  in the detailed  QCD sum rules analysis of  the photon light-cone distribution amplitudes \cite{PBall-photon}. \\

The existing  values of the vacuum susceptibilities $\chi$, $\kappa$ and $\xi$ are quite different from different determinations,
the old values $ \chi({\rm 1\, GeV})=-(4.4\pm 0.4)\,\rm{GeV}^{-2}$ \cite{Balitsky-chi-1985} or $ -(5.7\pm 0.6)\,\rm{GeV}^{-2}$ \cite{Belyaev-chi-1984} determined  in the  QCD sum rules combined with the vector meson dominance, while the most recent value $ \chi({\rm 1\, GeV})=-(2.85\pm 0.50)\,\rm{GeV}^{-2}$ determined  in the light-cone QCD sum rules for the radiative heavy meson decays \cite{Rohrwild-chi}.
 The old values $ \chi=-6\,\rm{GeV}^{-2}$, $\kappa=-0.75$, $\xi=1.5$
 at the energy scale  $\mu=1\, \rm{GeV}$ determined in the QCD sum rules for the octet and decuplet  baryon magnetic moments \cite{Magnetic-Lee-Formula} are also different from the parameters (I) determined in analysis of the magnetic moments of the $p$, $n$ and $\Lambda$  \cite{Magnetic-Pasupathy-Formula-Value,Magnetic-Chiu-Value}.  The most popularly  used values are the parameters (II) \cite{PBall-photon}, which are consistent  with the most recent values determined in Ref.\cite{Rohrwild-chi}. In the parameters (I), we choose the value $ \chi({\rm 1\, GeV})=-3\,\rm{GeV}^{-2}$ \cite{Magnetic-Pasupathy-Formula-Value,Magnetic-Chiu-Value} instead of $ \chi({\rm 1\, GeV})=-6\,\rm{GeV}^{-2}$ \cite{Magnetic-Lee-Formula} according to the most recent value $ \chi({\rm 1\, GeV})=-(2.85\pm 0.50)\,\rm{GeV}^{-2}$ \cite{Rohrwild-chi}.

  We take into account
the energy-scale dependence of  the input parameters from the renormalization group equation,
\begin{eqnarray}
\langle\bar{q}q \rangle(\mu)&=&\langle\bar{q}q \rangle(Q)\left[\frac{\alpha_{s}(Q)}{\alpha_{s}(\mu)}\right]^{\frac{12}{25}}\, ,\nonumber\\
\langle\bar{q}g_s \sigma Gq \rangle(\mu)&=&\langle\bar{q}g_s \sigma Gq \rangle(Q)\left[\frac{\alpha_{s}(Q)}{\alpha_{s}(\mu)}\right]^{\frac{2}{25}}\, , \nonumber\\
\chi(\mu)&=&\chi(Q)\left[\frac{\alpha_{s}(\mu)}{\alpha_{s}(Q)}\right]^{\frac{16}{25}} \, ,\nonumber\\
\kappa(\mu)&=&\kappa(Q)\left[\frac{\alpha_{s}(\mu)}{\alpha_{s}(Q)}\right]^{\frac{36}{25}} \, ,\nonumber\\
\xi(\mu)&=&\xi(Q)\left[\frac{\alpha_{s}(\mu)}{\alpha_{s}(Q)}\right]^{\frac{36}{25}} \, ,\nonumber\\
m_c(\mu)&=&m_c(m_c)\left[\frac{\alpha_{s}(\mu)}{\alpha_{s}(m_c)}\right]^{\frac{12}{25}} \, ,\nonumber\\
\alpha_s(\mu)&=&\frac{1}{b_0t}\left[1-\frac{b_1}{b_0^2}\frac{\log t}{t} +\frac{b_1^2(\log^2{t}-\log{t}-1)+b_0b_2}{b_0^4t^2}\right]\, ,
\end{eqnarray}
  where $t=\log \frac{\mu^2}{\Lambda^2}$, $b_0=\frac{33-2n_f}{12\pi}$, $b_1=\frac{153-19n_f}{24\pi^2}$, $b_2=\frac{2857-\frac{5033}{9}n_f+\frac{325}{27}n_f^2}{128\pi^3}$,  $\Lambda=210\,\rm{MeV}$, $292\,\rm{MeV}$  and  $332\,\rm{MeV}$ for the flavors  $n_f=5$, $4$ and $3$, respectively  \cite{PDG,PBall-photon,Narison-mix,Narison-Book}, and evolve all the input parameters to the optimal energy scale  $\mu=1.4\,\rm{GeV}$ to extract the form-factor $G_2(0)$ \cite{WangHuangTao-3900,WangZc3900Decay,Wang-1601}.

The hadronic parameters are taken as   $\sqrt{s^0_{Z}}=4.4\,\rm{GeV}$,  $M_{Z}=3.899\,\rm{GeV}$,   $\lambda_{Z}=2.1\times 10^{-2}\,\rm{GeV}^5$ \cite{WangHuangTao-3900,Wang-1601}.  In the  scenario of tetraquark  states, the QCD sum rules indicate that the $Z_c(3900)$ and $Z(4430)$ can be tentatively assigned to be the ground state and the first radial excited state of the axialvector tetraquark states, respectively \cite{Wang4430}. We choose the value $\sqrt{s^0_{Z}}=4.4\,\rm{GeV}$ to avoid contamination of the $Z(4430)$ and reproduce the experimental value  of the mass $M_{Z}=3.899\,\rm{GeV}$ from the QCD sum rules. The unknown parameter is fitted to be  $ C=0.97\,\rm{GeV}^{-2}$ and $ 0.94\,\rm{GeV}^{-2}$ for the parameters (I) and  (II) respectively to obtain  platforms  in the Borel window $T^2=(2.2-2.8)\,\rm{GeV}^2$.

\begin{figure}
 \centering
  \includegraphics[totalheight=5cm,width=7cm]{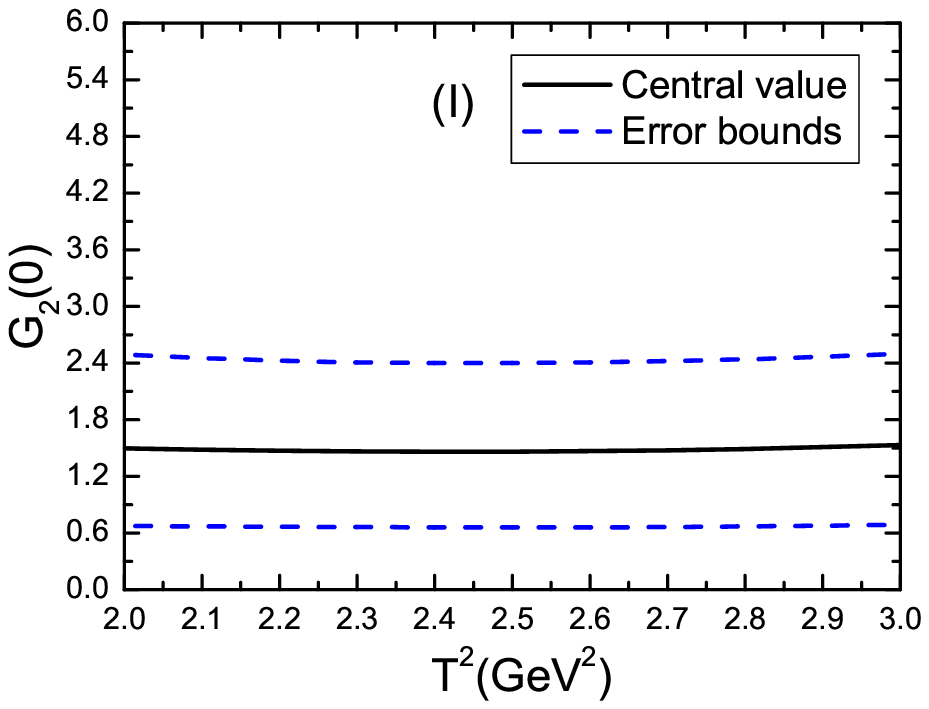}
  \includegraphics[totalheight=5cm,width=7cm]{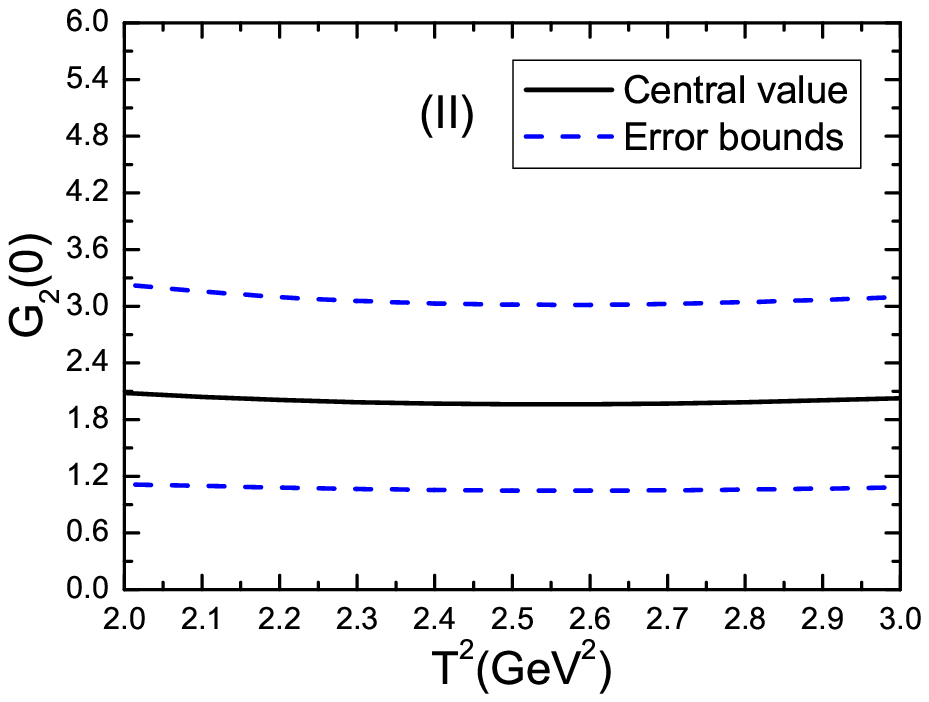}
         \caption{ The form-factor $G_2(0)$   with variation of the  Borel parameter $T^2$, where the (I) and (II) denote the parameters (I) and parameters (II), respectively.  }
\end{figure}

We take  into account  uncertainties of the input parameters,
and obtain the value of the form-factor $G_2(0)$ therefore the magnetic moment $\mu_Z$ of
 the    $Z_{c}(3900)$, which is shown explicitly in Fig.3,
\begin{eqnarray}
|\mu_{Z}|&=& G_2(0) \,\,\frac{e}{2M_{Z}} \nonumber\\
&=&1.46^{+0.98}_{-0.80} \,\,\frac{e}{2M_{Z}}   \nonumber\\
&=&0.35^{+0.24}_{-0.19}\,\mu_{N} \,\,\,\,\, {\rm for} \,\,\, {\rm parameters}\,\,\, {\rm (I)}\, , \\
&=&1.96^{+1.14}_{-0.91} \,\,\frac{e}{2M_{Z}}   \nonumber\\
&=&0.47^{+0.27}_{-0.22}\,\mu_{N}\,\,\,\,\, {\rm for} \,\,\, {\rm parameters}\,\,\, {\rm (II)}\, ,
\end{eqnarray}
where the $\mu_N$ is the nucleon magneton. The present prediction can be confronted to the experimental data in the future.

As the vacuum susceptibilities  $\kappa$ and $\xi$ are less well studied compared to the vacuum susceptibility $\chi$, in Fig.4, we plot the value of the $G_2(0)$ with variation of the parameter $y=-\kappa({\rm 1\,GeV})=\xi({\rm 1\,GeV})/2$ for the parameters $T^2=2.5\,\rm{GeV}^2$, $ \chi({\rm 1\,GeV})=-3.15\,\rm{GeV}^{-2}$ and $ C=0.94\,\rm{GeV}^{-2}$. From the figure, we can see that the value of the $G_2(0)$ decreases  monotonously with increase of the absolute values of the $\kappa$ and $\xi$. Once the value of the magnetic moment $\mu_Z$ is precisely measured, we can obtain a powerful  constraint on the values of the $\kappa$ and $\xi$.

\begin{figure}
 \centering
  \includegraphics[totalheight=6cm,width=9cm]{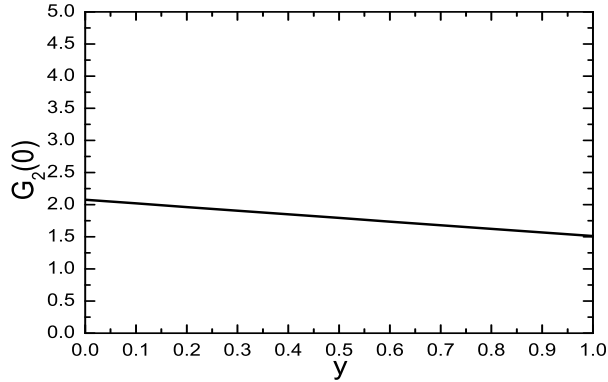}
           \caption{ The form-factor $G_2(0)$   with variation of the  parameter $y=-\kappa({\rm 1\,GeV})=\xi({\rm 1\,GeV})/2$.  }
\end{figure}

In Ref.\cite{LCSR-M-3900},  Ozdem and Azizi obtain the magnetic moment $\mu_{Z}=0.67 \pm 0.32\,\mu_N$ from the light-cone QCD sum rules, where the $\mu_{Z}$ decreases
 monotonously  with increase of the Borel parameter, the Borel window is not flat enough.  Compared to the QCD sum rules in the external electromagnetic field, the light-cone QCD sum rules have more parameters with limited precision. Furthermore, there exists no experimental data on the electromagnetic multipole moments
of the exotic particles $X$, $Y$ and $Z$, so theoretical studies  play an important role, more theoretical works are still needed. For example, the magnetic moment of the $Z_c(3900)$ as a molecular state is still not studied with the QCD sum rules or light-cone QCD sum rules, a comparison of  the two possible assignments is not possible at the present time.
We can diagnose  the nature of the $Z_c(3900)$ as a genuine resonance or
 anomalous triangle singularity in the photoproduction process \cite{Liu-3900}. Experimentally, the COMPASS collaboration  observed the $X(3872)$ in the subprocess
 $\gamma^*N\to N^\prime X(3872)\pi^\pm$ in the process $\mu^+N\to \mu^+ N^\prime X(3872)\pi^\pm\to\mu^+ N^\prime \pi^+\pi^- \pi^\pm$ \cite{COMPASS-3872}, but have not observed the $Z_c^\pm(3900)$ in the subprocess
 $\gamma^*N\to N^\prime Z_c(3900)^\pm$ in the process $\mu^+N\to \mu^+ N^\prime Z_c(3900)^\pm\to\mu^+ N^\prime J/\psi \pi^\pm$ yet \cite{COMPASS-3900}. We can study
 the photon associated  production  $\gamma^*N\to N^\prime Z_c(3900)^\pm \gamma$ or other processes with the final states $Z_c^\pm(3900)\gamma$ to study the electromagnetic multipole moments of the $Z_c^\pm(3900)$, although it is a difficult work.

\section{Conclusion}
In this article, we tentatively assign the $Z_c^\pm(3900)$ to be  the diquark-antidiquark  type axialvector  tetraquark  state,   study its magnetic moment with the QCD sum rules in the external weak electromagnetic field by carrying out the operator product expansion up to the vacuum condensates of dimension 8, and neglect the tiny contributions of the gluon condensate.     We pay  special attention  to matching the hadron side of the correlation function with the QCD side of the correlation function to obtain solid duality, the routine can be applied to study other electromagnetic properties of the  exotic particles $X$, $Y$ and $Z$  directly. Finally, we obtain the magnetic moment of the $Z_c(3900)$, which can be confronted to the experimental data in the future and shed light on the nature of the $Z_c(3900)$.

\section*{Acknowledgements}
This  work is supported by National Natural Science Foundation, Grant Number  11775079.


\begin{thebibliography}{99}

\bibitem{BES3900}   M. Ablikim et al, Phys. Rev. Lett. {\bf 110} (2013) 252001.

\bibitem{Belle3900} Z. Q. Liu   et al, Phys. Rev. Lett. {\bf 110} (2013) 252002.

\bibitem{CLEO3900} T. Xiao, S. Dobbs, A. Tomaradze and K. K. Seth, Phys. Lett. {\bf B727} (2013) 366.


\bibitem{JP-BES-Zc3900}  M. Ablikim  et al,  Phys. Rev. Lett. {\bf 119} (2017)  072001.

\bibitem{Maiani1303} R. Faccini, L. Maiani, F. Piccinini, A. Pilloni, A. D. Polosa and V. Riquer, Phys. Rev. {\bf D87} (2013) 111102(R).

\bibitem{Tetraquark3900} M. Karliner and S. Nussinov, JHEP {\bf 1307} (2013) 153;
N. Mahajan, arXiv:1304.1301;
E. Braaten, Phys. Rev. Lett. {\bf 111} (2013) 162003;
C. F. Qiao and L. Tang, Eur. Phys. J. {\bf C74} (2014) 3122.

\bibitem{WangHuangTao-3900} Z. G. Wang and T. Huang,  Phys. Rev. {\bf D89} (2014) 054019.

\bibitem{Nielsen3900} J. M. Dias, F. S. Navarra, M. Nielsen and C. M. Zanetti, Phys. Rev. {\bf D88} (2013) 016004.

\bibitem{Molecular3900} F. K. Guo, C. Hidalgo-Duque, J. Nieves and M. P. Valderrama, Phys. Rev. {\bf D88} (2013) 054007;
J. R. Zhang, Phys. Rev. {\bf D87} (2013) 116004;
Y. Dong, A. Faessler, T. Gutsche and V. E. Lyubovitskij, Phys. Rev. {\bf D88} (2013) 014030;
H. W. Ke, Z. T. Wei and X. Q. Li,  Eur. Phys. J. {\bf C73} (2013) 2561;
S. Prelovsek and L. Leskovec, Phys. Lett. {\bf B727} (2013) 172;
C. Y. Cui, Y. L. Liu, W. B. Chen and M. Q. Huang,  J. Phys. {\bf G41} (2014) 075003.

\bibitem{WangMolecule-3900} Z. G. Wang and T. Huang, Eur. Phys. J. {\bf C74} (2014) 2891;
Z. G. Wang, Eur. Phys. J. {\bf C74} (2014)  2963.

\bibitem{SVZ79} M. A. Shifman, A. I. Vainshtein and V. I. Zakharov, Nucl. Phys. {\bf B147} (1979) 385; Nucl. Phys. {\bf B147} (1979) 448.

\bibitem{Reinders85} L. J. Reinders, H. Rubinstein and S. Yazaki, Phys. Rept. {\bf 127} (1985) 1.

\bibitem{Azizi-Zc3900-decay} S. S. Agaev, K. Azizi and H. Sundu, Phys. Rev. {\bf D93} (2016)  074002.

\bibitem{WangZc3900Decay} Z. G. Wang and J. X. Zhang, Eur. Phys. J. {\bf C78} (2018)  14.

\bibitem{LCSR-M-3900} U. Ozdem and K. Azizi, Phys. Rev. {\bf D96} (2017) 074030.

\bibitem{LCSR-MM-XYZ}  A. K. Agamaliev, T. M. Aliev and M. Savci, Phys. Rev. {\bf D95} (2017)  036015;
U. Ozdem and K. Azizi, Phys. Rev. {\bf D97} (2018)  014010.


\bibitem{Ioffe1984} B. L. Ioffe and A. V. Smilga, Nucl. Phys. {\bf B232} (1984) 109.

\bibitem{Balitsky1983}  I. I. Balitsky and A. V. Yung, Phys. Lett. {\bf 129B} (1983) 328.



\bibitem{Magnetic-Chiu} C. B. Chiu, J. Pasupathy and S. L. Wilson, Phys. Rev. {\bf D33} (1986) 1961.

\bibitem{Magnetic-Pasupathy-Formula-Value} J. Pasupathy, J. P. Singh, C. B. Chiu and S. L. Wilson, Phys. Rev. {\bf D36} (1987) 1442.

\bibitem{Magnetic-Chiu-Value} C. B. Chiu, J. Pasupathy and S. L. Wilson, Phys. Rev. {\bf D36} (1987) 1451.

\bibitem{Magnetic-Lee-Formula} F. X. Lee, Phys. Rev. {\bf D57} (1998) 1801;
L. Wang and F. X. Lee, Phys. Rev. {\bf D78} (2008) 013003.


\bibitem{Brodsky1992}  S. J. Brodsky and J. R. Hiller, Phys. Rev. {\bf D46} (1992) 2141.

\bibitem{PDG}   C. Patrignani et al, Chin. Phys. {\bf C40} (2016)  100001.

\bibitem{Wang4430} Z. G. Wang,  Commun. Theor. Phys. {\bf 63} (2015) 325.

\bibitem{Wang-EPJC-Scalar-meson} Z. G. Wang,  Eur. Phys. J. {\bf C76} (2016) 427.

\bibitem{Wang-Zc4200} Z. G. Wang, Int. J. Mod. Phys. {\bf A30} (2015) 1550168.

\bibitem{Pascual-1984} P. Pascual and R. Tarrach, ``QCD: Renormalization for the practitioner", Springer Berlin Heidelberg (1984).


\bibitem{Colangelo-Review}  P. Colangelo and A. Khodjamirian, hep-ph/0010175.



\bibitem{PBall-photon} P. Ball, V. M. Braun and N. Kivel, Nucl. Phys. {\bf B649} (2003) 263.

\bibitem{Rohrwild-chi} J. Rohrwild,  JHEP {\bf 0709} (2007) 073.

\bibitem{Balitsky-chi-1985} I. I. Balitsky, A. V. Kolesnichenko and A. V. Yung, Sov. J. Nucl. Phys. {\bf 41}  (1985)  178.

\bibitem{Belyaev-chi-1984} V. M. Belyaev and Y. I. Kogan, Yad. Fiz. {\bf 40} (1984) 1035.


\bibitem{Narison-mix} S. Narison and R. Tarrach, Phys. Lett. {\bf 125 B} (1983) 217.

\bibitem{Narison-Book} S. Narison, ``QCD as a theory of hadrons from partons to confinement", Camb. Monogr. Part. Phys. Nucl. Phys. Cosmol. {\bf 17} (2007) 1.


\bibitem{Wang-1601} Z. G. Wang,  Eur. Phys. J. {\bf C76} (2016)  387.


\bibitem{Liu-3900} Q. Y. Lin, X. Liu and H. S. Xu,  Phys. Rev. {\bf D88} (2013)  114009.

\bibitem{COMPASS-3872} M. Aghasyan et al, arXiv:1707.01796.

\bibitem{COMPASS-3900} C. Adolph et al, Phys. Lett. {\bf B742} (2015) 330.

\end{thebibliography}
\end{document}